\documentclass[aps,pra,twocolumn,showpacs]{revtex4-1}
\usepackage{eurosym}
\usepackage{amsmath, amsthm, amssymb}
\usepackage{graphicx}
\usepackage{color}
\usepackage{times}
\usepackage{epstopdf}
\usepackage[usenames,dvipsnames,svgnames,table]{xcolor}
\usepackage{multirow}
\usepackage{natbib}
\usepackage[colorlinks=true, citecolor=blue, linkcolor=blue, urlcolor=blue]{hyperref}

\begin{document}

\title{Superfluid-Mott insulator transition in spin-orbit coupled Bose-Hubbard Model}
\author{A. T. Bolukbasi} \author{M. Iskin}
\affiliation{Department of Physics, Ko\c c University, Rumelifeneri Yolu, 34450 Sar{\i}yer, Istanbul, Turkey.}
\date{\today }

\begin{abstract}

We consider a square optical lattice in two dimensions and study the effects of
both the strength and symmetry of spin-orbit-coupling (SOC) and Zeeman field
on the ground-state, i.e., Mott insulator (MI) and superfluid (SF), phases and
phase diagram, i.e., MI-SF phase transition boundary, of the two-component
Bose-Hubbard model. In particular, based on a variational Gutzwiller ansatz,
our numerical calculations show that the spin-orbit coupled SF phase is a nonuniform
(twisted) one with its phase (but not the magnitude) of the order parameter modulating
from site to site. Fully analytical insights into the numerical results are also given.

\end{abstract}

\maketitle

\section{INTRODUCTION}
\label{sec:intro}

Ultracold atoms have proved to be exceptional many-body quantum
systems, thanks especially to their tunable single-particle potentials and
multi-particle interactions. The experimental ability
in controlling the parameters of the atomic Hamiltonian allow one to
simulate and study some of the fundamental aspects of condensed-matter
systems, including Bose-Einstein condensation (BEC), bosonic superfluidity,
quantum magnetism, many-body spin dynamics,
Bardeen-Cooper-Schrieffer (BCS) superfluidity, BEC-BCS crossover,
etc.~\cite{lewenstein07, bloch08, giorgini08}.
In addition, by loading cold atoms into the periodic optical lattice potentials,
which are formed by interfering counter-propagating laser beams,
it has also been possible to realize Hubbard-type lattice models and
study strongly-correlated quantum phenomena~\cite{lewenstein07, bloch08}.
In particular, bosonic atoms in an optical lattice, whose low-energy
dynamics is well-captured by the Bose-Hubbard model~\cite{fisher89},
provide an ideal platform for the observation of Mott insulator (MI)
and superfluid (SF) phases as well as the MI-SF phase transition between
the two~\cite{lewenstein07, bloch08}.

Meanwhile, recent discoveries of topological insulators~\cite{hasan10},
topological superconductors~\cite{qi11} and quantum spin-Hall
effect~\cite{maciejko11} have put topological phases
of matter on the spotlight. It turns out that the interaction between the
quantum particle's spin and its center-of-mass motion (momentum),
i.e., spin-orbit coupling (SOC), is at the heart of all of these modern
condensed-matter phenomena, and creation and manipulation of a
similar (if not identical) effect has been an intriguing possibility for the
cold-atom community. However, since quantum gases are charge neutral,
they do not directly couple to electromagnetic fields, and this prevented SOC
studies in atomic systems until to the advent of artificial
gauge fields~\cite{dalibard11,galitski13}.
By coupling the internal states of atoms to their momentum via Raman
dressing of atomic hyperfine states with near-resonant laser beams,
it has recently been possible to engineer
atomic systems with Abelian gauge fields.
For instance, while there are many proposals for implementing
atomic gases with various non-Abelian gauge fields that may give
rise to Rashba, Dresselhaus and Weyl SOCs,
several experimental groups have so far achieved only a particular
form of an Abelian gauge field that may be characterized as an equal
Rashba and Dresselhaus (ERD)
SOC~\cite{lin11, chen12, wang12, cheuk12, hamner13, fu13, williams13}.
Note that a very recent proposal to realize SOC in optical lattices
does not rely on laser light to couple different spin states~\cite{colin13}.
These experiments naturally lead the way to numerous other works
on spin-orbit-coupled atomic systems, e.g., topological superfluid
phases of matter, bringing once again the condensed-matter
and atomic-physics communities together.

Motivated by these developments, here we consider a square lattice in two
dimensions and study the effects of both the strength and symmetry of
SOC and Zeeman field on the ground-state phases and phase diagram of the
two-component Bose-Hubbard model. In particular, based on a
variational Gutzwiller ansatz, we analyze the competition between the
interaction, tunneling, Rashba and ERD SOCs, and out-of- and in-plane
Zeeman fields on the MI-SF phase transition boundary and the nature
of the SF phase nearby.
In addition to the phase diagrams, one of our main results is as follows:
Gutzwiller calculations show that while the magnitudes of the order
parameters are uniform across the entire lattice, their phases may
vary from site to site due to SOC, and therefore, the SF phase is a
nonuniform one. We give a complete account and intuitive understanding
of this SOC induced nonuniform-SF phase and its resultant phase
patterns, by supporting our numerical calculations with fully
analytical insights.

The rest of the paper is organized as follows. In Sec.~\ref{sec:tcbhm}, we
introduce the spin-orbit coupled two-component Bose-Hubbard model, and
derive the self-consistency (total average number, polarization and SF order
parameter) equations using a variational Gutzwiller ansatz. Our numerical
results for the ground-state MI, uniform SF and nonuniform SF phases, and
the MI-SF phase transition boundary are presented in Sec.~\ref{sec:nr} as
functions of the strength and the symmetry of the SOC and Zeeman field.
The paper is concluded with a brief summary of our results and an outlook
in Sec.~\ref{sec:conc}.

\section{Two-Component Bose-Hubbard Model}
\label{sec:tcbhm}

It has long been established that the low-energy dynamics of quantum gases
loaded into the periodic optical lattice potentials are well-described by
Hubbard-type tight-binding lattice models~\cite{lewenstein07, bloch08}. In particular,
the simplest Bose-Hubbard model~\cite{fisher89}, which takes into account
the chemical potential and nearest-neighbor tunneling of atoms, and
short-range (on-site) repulsive interparticle interactions, has proved to be
quite successful in describing some of the cold-atom experiments where MI
and uniform SF phases as well as the MI-SF phase transition were
observed~\cite{bloch08}. This success generated an enormous interest in this
topic, and many extensions of Hubbard-type models have not only been
proposed but also realized in the recent literature, including different
lattice geometries, longer-ranged tunnelings and interactions, multiple
components, gauge fields, etc. A number of theoretical
methods have been developed to tackle these lattice models, and among
those the validity and limitations of the variational Gutzwiller
ansatz~\cite{sheshadri93}, decoupling mean-field theory~\cite{oosten01},
strong-coupling expansion~\cite{freericks96, freericks09}, and quantum monte
carlo~\cite{freericks09, pollet13} approaches are well understood.

In this context, the two-component Bose-Hubbard
model~\cite{kuklov03, altman03, kuklov04, isacsson05, arguelles07, hu09, iskin10, ozaki12}
was introduced about a decade ago to describe cold-atom experiments
involving two types of bosons, in which the two components may
correspond to different hyperfine states of a particular atom
or different species of atoms. In addition to the
phases that are similar in many ways to the MI and SF phases of the
single-component model, these works proposed that paired-SF, counterflow-SF,
density-wave insulator and supersolid phases may be created with the
experimental realization of the two-component model. These possibilities
already motivated a number of experimental studies on two-component
systems~\cite{thalhammer08, widera08, weld09, gadway10},
opening up a new frontier waiting to be explored in the near future.

In addition, excited by the recent realization of spin-orbit-coupled atomic
BEC~\cite{lin11, chen12, colin13}, there has been growing interest in
studying the effects of SOC on the two-component
model~\cite{cole12, radic12, cai12, zhou13, zhang03, zhao03, mandal12, wong13, grass13, sakaguchi13}.
For instance, it has been proposed that the SOC gives rise to rich
phase diagrams which exhibit spin textures in the form of spin spirals
and vortex and Skyrmion crystals within the MI
phase~\cite{cole12, radic12, cai12, zhou13, zhang03},
and also a nonuniform twisted SF phase~\cite{mandal12}. Our main goal here
is to provide a complete analysis of this SOC induced nonuniform-SF
phase as well as the MI-SF phase transition boundary.

\subsection{Hamiltonian: SOC and Zeeman fields}
\label{sec:bhm}

In this paper, we consider a square lattice in two dimensions and study the
effects of both the strength and symmetry of SOC and Zeeman field on the
ground-state phases and phase diagram of the two-component
Bose-Hubbard model. The Hamiltonian of such systems may be written as
\begin{eqnarray}
H &=&\sum_{j\alpha }\left[ \frac{U_{\alpha \alpha }}{2}\hat{n}_{j\alpha }(
\hat{n}_{j\alpha }-1)-\mu _{\alpha }\hat{n}_{j\alpha }\right] +U_{\uparrow
\downarrow }\sum_{j}\hat{n}_{j\uparrow }\hat{n}_{j\downarrow }  \notag \\
&&-\sum_{\langle j,k \rangle}\left( \hat{\Psi}_{j}^{\dagger }T^{jk}\hat{\Psi}
_{k}+H.c.\right) -h_{y}\sum_{j}\hat{\Psi}_{j}^{\dagger }\sigma _{y}\hat{\Psi}
_{j},  \label{eqn:ham}
\end{eqnarray}
where $\alpha \equiv (\uparrow, \downarrow )$ denotes the two types of
bosons, $U_{\alpha \alpha} \ge 0$ is the intra-component and $%
U_{\uparrow \downarrow} \ge 0$ is the inter-component interaction with $%
U_{\uparrow\downarrow}^2<U_{\uparrow\uparrow} U_{\downarrow\downarrow}$ to
prevent phase separation, and the operator $\hat{n} _{j\alpha }=\hat{a}_{j\alpha
}^{\dagger }\hat{ a}_{j\alpha }$ counts the local number of $\alpha$ bosons
at site $j$. Here, the operator $\hat{a} _{j\alpha }^{\dagger } (\hat{a%
}_{j\alpha })$ creates (annihilates) an $\alpha $ boson at site $j$. The
chemical potential $\mu _{\alpha }$ already includes the out-of-plane $h_z$
component of the Zeeman field such that $\mu _{\uparrow }=\mu +h_{z}$
and $\mu _{\downarrow }=\mu-h_{z}$. In the second line, $\langle j,k \rangle$ sums over
the nearest-neighbor sites, where the operator
$
\hat{\Psi}_{j}=
\begin{pmatrix}
\hat{a}_{j\uparrow } & \hat{a}_{j\downarrow }%
\end{pmatrix}^{T}
$
denotes the boson operators collectively, $H.c.$ is the Hermitian
conjugate, $h_y$ is the in-plane component of the Zeeman field, and $%
\sigma_y$ is the Pauli spin matrix. In Eq.~(\ref{eqn:ham}), we set the
in-plane $h_x$ component of the Zeeman field to 0 without loosing
generality.

In Eq.~(\ref{eqn:ham}), the spin matrices $T$ include both the
spin-preserving and spin-flipping nearest-neighbor tunnelings, and they can
be compactly written as $T^{j,j\pm \hat{x}}=t\sigma _{0}\pm i\gamma
_{x}\sigma _{y} $ for hoppings in the $\pm x$-direction and $\ T^{j,j\pm
\hat{y}}=t\sigma _{0}\mp i\gamma _{y}\sigma_{x}$ for hoppings in the
$\pm y$-direction, where $t$ is the strength of the usual single-particle
tunneling with $\sigma_0$ the identity matrix, and the parameters
$\gamma_x \ge 0$ and $\gamma_y \ge 0$ characterize the strength
and symmetry of the SOC. These spin matrices can be derived from
a non-Abelian gauge field $\vec{A}=(\beta _{x}\sigma
_{y},-\beta _{y}\sigma _{x},0)$, where $\beta_x$ and $\beta_y$ are constants
in space, using the Peierl's substitution. This leads to $%
T^{jk}=t_{0}e^{i\int_{j}^{k}\vec{A}\cdot d\vec{r}}$, such that $T^{j,j\pm
\hat{x}} = t_{0}\cos \beta _{x}\sigma_{0} \pm it_{0}\sin \beta _{x} \sigma
_{y} $ for tunnelings in the $\pm x$-direction and $T^{j,j\pm \hat{y}
}=t_{0}\cos \beta _{y} \sigma_{0} \mp i t_{0}\sin \beta _{y} \sigma _{x} $
for tunnelings in the $\pm y$-direction. Thus, our model parameters in
Eq.~(\ref{eqn:ham}) are related to the parameters of the gauge field $\vec{A}
$ via $\gamma_x= t \tan \beta_{x}$ and $\gamma_y= t \tan \beta_{y}$. Note
that the ratio of $\gamma_x$ and $\gamma_y$ determines the symmetry of the
SOC, and we compare and discuss three distinct limits throughout this paper:
(i) Rashba SOC where $\gamma _{x} = \gamma _{y} = \gamma_R \ne 0$, (ii) ERD$%
_x$ SOC where $\gamma _{x} \ne 0$ and $\gamma _{y}=0$, and (iii)
ERD$_{y}$ SOC where $\gamma _{x}=0$ and $\gamma _{y} \ne 0$.

It is very difficult to obtain the exact solutions for the model Hamiltonian
given in Eq.~(\ref{eqn:ham}) even in the absence of inter-component
interaction, SOC and Zeeman field. Therefore, hoping to produce
qualitatively accurate ground-state phases and phase diagrams,
next we propose a properly generalized variational Gutzwiller ansatz
for our model.

\subsection{Variational Gutzwiller ansatz}
\label{sec:GA}

The variational Gutzwiller ansatz for the approximate many-body
wave function $|\psi \rangle$ is a product state that is formed by
multiplying local ground states $|\psi^j\rangle$ of the entire lattice,
i.e. $|\psi \rangle = \prod_{j} |\psi^j \rangle$, and thus,
it neglects the off-site correlations by construction.
The simpler versions of this ansatz have been frequently
used in the literature to approximate the ground-state wave functions of
Bose-Hubbard type Hamiltonians at zero temperature. In
the single-component case, since the ansatz reproduces (by construction) the
exact ground states of the system in the extremely-strong (i.e., deep in the
MI phase) and extremely-weak (i.e., deep in the SF phase) interaction
limits, it naturally works qualitatively well in between for the MI-SF phase
transition boundary. Earlier works also showed that the results obtained
from this ansatz precisely matches those of the mean-field decoupling
approximation for the MI-SF phase transition boundary, and therefore, the
level of approximation (i.e., negligence of the off-site correlations)
are considered to be exactly equivalent in both
methods~\cite{fisher89, sheshadri93, oosten01}.

The generalized Gutzwiller wave function for the model Hamiltonian given in
Eq.~(\ref{eqn:ham}) can be written as
\begin{equation}
|\psi \rangle =\prod_{j}\left( \sum_{l_{\uparrow }l_{\downarrow
}}f_{l_{\uparrow }l_{\downarrow }}^{j}|l_{\uparrow },l_{\downarrow }\rangle
_{j}\right),  \label{eqn:GA}
\end{equation}
where the complex variational parameter $f_{l_{\uparrow }l_{\downarrow
}}^{j} $ determines the probability amplitude of the occupation of the Fock
state $|l_{\uparrow },l_{\downarrow }\rangle _{j}$ at site $j$. Here, the
local Fock state is characterized by the occupation of ($l_{\uparrow},
l_{\downarrow}$) bosons from each type, where $l_{\alpha}=0,1,...,l_{\max }$
and $l_{\max }$ is the maximum number of $\alpha$ bosons allowed in the
numerics (to be specified in Sec.~\ref{sec:nr}). The normalization of the
wave function $\langle \psi |\psi \rangle =1$ requires $\sum_{l_{\uparrow }
l_{\downarrow}} |f_{l_{\uparrow }l_{\downarrow }}^{j}|^{2}=1$ for each site $j$.

Given the ground-state ansatz, it is a straightforward task to calculate any
of the desired observables. For instance, we are interested in the average
number of local $\alpha$ bosons $N_{j\alpha }=\langle \psi |\hat{n}_{j\alpha
}|\psi \rangle$ and the projections of average local polarizations $%
P_{jq}=\langle \psi |\hat{\Psi}_{j}^{\dagger }\sigma _{q}\hat{\Psi}_{j}|\psi
\rangle$ along the $q\equiv (x,y,z)$-direction. Using Eq.~(\ref{eqn:GA}),
and after some algebra, we obtain
\begin{eqnarray}
N_{j} &=& \sum_{l_{\uparrow }l_{\downarrow }}\left( |f_{l_{\uparrow
}l_{\downarrow }}^{j}|^{2}l_{\uparrow }+|f_{l_{\uparrow }l_{\downarrow
}}^{j}|^{2}l_{\downarrow }\right),
\label{eqn:ntot} \\
P_{jz} &=& \sum_{l_{\uparrow }l_{\downarrow }}\left( |f_{l_{\uparrow
}l_{\downarrow }}^{j}|^{2}l_{\uparrow }-|f_{l_{\uparrow }l_{\downarrow
}}^{j}|^{2}l_{\downarrow }\right),
\label{eqn:pz} \\
P_{jy} &=&2 \text{Im} \sum_{l_{\uparrow }l_{\downarrow }} f_{l_{\uparrow
}l_{\downarrow }}^{j\ast }f_{l_{\uparrow }-1,l_{\downarrow }+1}^{j} \sqrt{%
l_{\uparrow }(l_{\downarrow }+1)},
\label{eqn:py}
\end{eqnarray}
where $N_j = N_{j\uparrow} + N_{j\downarrow}$ is the total average number of
bosons on site $j$ and $P_{jz} = N_{j\uparrow} - N_{j\downarrow}$.
Here, Im$[\cdots]$ is the imaginary part of $[\cdots]$, and the real part of
the same sum gives $P_{jx}$. Note that while the overall $x$-component of
the average polarization $\sum_j P_{jx} = 0$, since we already set $h_x = 0$
in Eq.~(\ref{eqn:ham}), SOC may still induce local $P_{jx} \ne 0$, causing
Skyrmion-like spin textures.
As discussed in Sec.~\ref{sec:nr}, all of our numerical calculations show
that average particle numbers are uniform across the entire lattice, and hence,
we also define $N = N_j$ and $N_{\alpha}=N_{j\alpha}$ for all $j$.

In order to distinguish the SF and non-SF (e.g., MI) ground states of the
system, the local average number and polarization
Eqs.~(\ref{eqn:ntot})-(\ref{eqn:py}) need to be solved self-consistently
with the local single-particle/single-hole SF order parameters $\Delta
_{j\alpha}=\langle \psi | \hat{a}_{j\alpha }|\psi \rangle$. Note that
exotic SF phases involving multi particle and/or hole excitations are
not accessible with this definition, and they are not of our main interest
in this work (see also Sec.~\ref{sec:conc}).
Using Eq.~(\ref{eqn:GA}), and after some algebra, we obtain
\begin{eqnarray}
\Delta _{j\uparrow } &=&\sum_{l_{\uparrow }l_{\downarrow }} f_{l_{\uparrow
}l_{\downarrow }}^{j\ast }f_{l_{\uparrow }+1,l_{\downarrow }}^{j}\sqrt{%
l_{\uparrow }+1},  \label{eqn:opup} \\
\Delta _{j\downarrow } &=&\sum_{l_{\uparrow }l_{\downarrow }} f_{l_{\uparrow
}l_{\downarrow }}^{j\ast }f_{l_{\uparrow },l_{\downarrow }+1}^{j}\sqrt{%
l_{\downarrow }+1},  \label{eqn:opdo}
\end{eqnarray}
which are complex numbers in general. As discussed in Sec.~\ref{sec:nr}, all
of our numerical results showed that while the magnitudes of these
parameters are uniform across the entire lattice, their phases are
nonuniform in general, i.e., $\theta _{j\alpha}=\arg(\Delta _{j\alpha})$
are not equal for all $j$. In this paper, we set the phase of the $\uparrow$
order parameter on some reference lattice site (which is labeled throughout
this paper as $j \equiv 0$) to 0, i.e., $\theta _{0\uparrow }=0$, and define all of the
remaining $\theta _{j\alpha}$ with respect to this reference site. Thus, in
Sec.~\ref{sec:nr}, we define $\Delta _{j\alpha} = \overline{\Delta}_\alpha
e^{i\theta_{j\alpha}}$, and distinguish the SF phases from the MI ones by
looking at whether the minimum energy configuration has $\overline{\Delta}
_\alpha \ne 0$ or 0. In addition, we distinguish the uniform-SF phase from
nonuniform-SF ones based on whether the minimum energy configuration
has a uniform $\theta_{j\alpha} = \theta_{\alpha}$ for all $j$ or not. Note that,
depending on the model parameters, we may have $\overline{\Delta}
_\alpha = 0$ and $\overline{\Delta}_{-\alpha} \ne 0$, so that the
ground-state is a mixture of $\alpha$-MI and $(-\alpha)$-SF, where $%
(-\uparrow) \equiv \downarrow$ and vice versa.

In the self-consistency Eqs.(\ref{eqn:ntot})-(\ref{eqn:opdo}), the set of
variational parameters $\{f_{l_{\uparrow }l_{\downarrow }}\}$ is determined
by minimizing the ground-state energy of the system. For this purpose, we
solve the Schr\"{o}dinger equation, i.e., $\langle \psi |H|\psi \rangle
=i\hbar \langle \psi |\partial |\psi \rangle /\partial \tau $, where we set $%
f_{l_{\uparrow }l_{\downarrow }}^{j}(\tau )=f_{l_{\uparrow }l_{\downarrow
}}^{j}e^{-iE_{0}\tau /\hbar }$ with $E_{0}$ the local ground-state energy of
the system and $\tau $ the time. Using Eq.~(\ref{eqn:GA}), and after some
algebra, we obtain
\begin{eqnarray}
&&E_{0}f_{l_{\uparrow }l_{\downarrow }}^{j}=f_{l_{\uparrow }l_{\downarrow
}}^{j}\left\{ U_{\uparrow \downarrow }l_{\uparrow }l_{\downarrow
}+\sum_{\alpha }\left[ \frac{U_{\alpha \alpha }}{2}l_{\alpha }(l_{\alpha
}-1)-\mu _{\alpha }l_{\alpha }\right] \right\}  \notag \\
&&-\sum_{\alpha ,k_{j}}\left[ \Delta _{k\alpha }\left( T_{\uparrow \alpha
}^{jk}\sqrt{l_{\uparrow }}f_{l_{\uparrow }-1,l_{\downarrow
}}^{j}+T_{\downarrow \alpha }^{jk}\sqrt{l_{\downarrow }}f_{l_{\uparrow
},l_{\downarrow }-1}^{j}\right) \right.  \notag \\
&&\left. +\Delta _{k\alpha }^{\ast }\left( T_{\uparrow \alpha }^{jk\ast }
\sqrt{l_{\uparrow }+1}f_{l_{\uparrow }+1,l_{\downarrow }}^{j}+T_{\downarrow
\alpha }^{jk\ast }\sqrt{l_{\downarrow }+1}f_{l_{\uparrow },l_{\downarrow
}+1}^{j}\right) \right]  \notag \\
&&+ih_{y}\left[ \sqrt{l_{\uparrow }(l_{\downarrow }+1)}f_{l_{\uparrow
}-1,l_{\downarrow }+1}^{j}-\sqrt{(l_{\uparrow }+1)l_{\downarrow }}
f_{l_{\uparrow }+1,l_{\downarrow }-1}^{j}\right] ,  \label{eqn:se}
\end{eqnarray}
where $k_{j}$ sums over the nearest-neighbors $k$ of site $j$. We note
that all of the tunneling and SOC terms vanish in the MI phase
when $\overline{\Delta }_{\alpha }=0$, and therefore, recently proposed
magnetic (spin-textured) MI phases~\cite{cole12, radic12, cai12, zhou13, zhang03}
are not accessible within our Gutzwiller ansatz. However, the method may still give
a quantitatively accurate description of the MI-SF phase transition boundary
as well as the nonuniform SF phases near this boundary. To understand the
competition between the interaction, tunneling, SOC and Zeeman field,
and the resultant MI and SF phases, let us first discuss the classical limit
and analyze the ground-state phase diagram of the system in the atomic limit.

\subsection{Atomic limit: MI phases}
\label{sec:al}

Setting $t = \gamma_x = \gamma_y = 0$ in the Hamiltonian decouples all of
the lattice sites from each other, and therefore, it is sufficient to
consider a single site to understand the resultant MI phases. First
of all, in contrast with the $h_y=0$ case where $N_\alpha$ is conserved for
both $\alpha$ bosons, only the total number $N = N_\uparrow + N_\downarrow$
of bosons is a good quantum number when $h_y \ne 0$. Thus, the MI lobes must
be labeled accordingly. Using Gutzwiller-like local ground states $|\psi
_{N}\rangle =\sum_{l_{\uparrow }l_{\downarrow }\ni
l_{\uparrow }+l_{\downarrow }=N}f_{l_{\uparrow }l_{\downarrow }}|l_{\uparrow
},l_{\downarrow }\rangle $, which can be shown to be exact for a given total
particle sector $N$, we can easily obtain the exact local ground-state
energy $\mathcal{E}_{N}=\langle \psi _{N}|H|\psi _{N}\rangle$ of the system
by minimizing $\mathcal{E}_N$ with respect to $f_{l_{\uparrow
}l_{\downarrow}}$.

For instance, $\mathcal{E}_{0}=0$ in the trivial case when $N=0$, and its
corresponding eigenstate is the vacuum state $|0,0\rangle$ with $f_{00} = 1$.
There are two energy eigenvalues when $N=1$, and $\mathcal{E}_{1}$ can be
written as
\begin{equation}
\mathcal{E}_{1}=\Phi _{1}^{^{\dagger }}
\begin{pmatrix}
-\mu _{\uparrow } & ih_{y} \\
-ih_{y} & -\mu _{\downarrow }%
\end{pmatrix}
\Phi _{1},  \label{en1}
\end{equation}
where $\Phi _{1}=
\begin{pmatrix}
f_{10} & f_{01}%
\end{pmatrix}%
^{T}. $ Likewise, there are three energy eigenvalues when $N=2$, and $%
\mathcal{E}_{2}$ can be written as
\begin{equation}
\mathcal{E}_{2}=\Phi _{2}^{^{\dagger }}\left(
\begin{array}{ccc}
-2\mu +U_{\uparrow \downarrow } & -i\sqrt{2}h_{y} & i\sqrt{2}h_{y} \\
i\sqrt{2}h_{y} & -2\mu _{\uparrow }+U_{\uparrow \uparrow } & 0 \\
-i\sqrt{2}h_{y} & 0 & -2\mu _{\downarrow }+U_{\downarrow \downarrow }%
\end{array}
\right) \Phi _{2},  \label{en2}
\end{equation}
where $\Phi _{2}=
\begin{pmatrix}
f_{11} & f_{20} & f_{02}%
\end{pmatrix}^{T}$.
All of the energy eigenvalues and their corresponding eigenstates
can be easily obtained by diagonalizing such matrices for any given $N$, and
$E_0$ corresponds to the minimal eigenvalue.

\begin{figure}[h]
\begin{center}
\includegraphics[width=220pt]{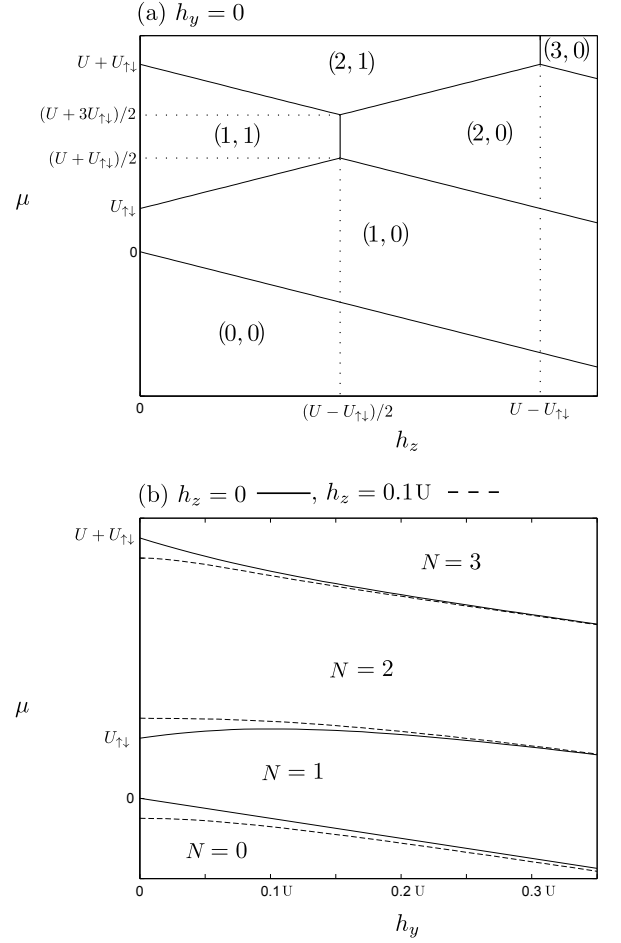}
\end{center}
\caption{
The atomic-limit ($t = \protect\gamma_x = \protect\gamma_y = 0$)
phase diagrams are shown as functions of (a) $\protect\mu$ and $h_{z}$ for $%
h_{y}=0$, and of (b) $\protect\mu$ and $h_{y}$ for $h_{z}=0$ (solid line)
and $h_{z}=0.1U$ (dashed line). The MI lobes are labeled by $(N_{\uparrow
},N_{\downarrow })$ in (a) and $N = N_\uparrow + N_\downarrow$ in (b). While
we set $U_{\uparrow\downarrow }=0.3U$ in these figures, they are
schematically correct as long as $0<U_{\uparrow\downarrow}<U=U_{\uparrow%
\uparrow}=U_{\downarrow\downarrow}$.
}
\label{fig:t_0}
\end{figure}

In Fig.~\ref{fig:t_0}, we present the atomic-limit phase diagrams as
functions of (a) $\mu$ and $h_{z}$ for $h_{y}=0$, and of (b) $\mu$ and $%
h_{y} $ for $h_{z}=0$ and $h_{z}=0.1U$. The MI lobes are naturally labeled
by $(N_{\uparrow },N_{\downarrow })$ in (a) and $N$ in (b) as explained
above. While we set $U_{\uparrow \downarrow }=0.3U$ in these figures, they
are schematically correct as long as $0<U_{\uparrow\downarrow}<U=U_{\uparrow
\uparrow}=U_{\downarrow\downarrow}$. When $h_y=0$, Fig.~\ref{fig:t_0}(a)
shows that the size of the $N=1$ lobe grows as $h_{z}$ increases toward $%
(U-U_{\uparrow \downarrow })/2$ and its size remains essentially unchanged
for $h_{z}>(U-U_{\uparrow \downarrow })/2$. This is in contrast with the $%
N=2 $ MI lobe, the size of which shrinks as $h_{z}$ increases toward $%
(U-U_{\uparrow \downarrow })/2$, followed by an increase between $%
(U-U_{\uparrow \downarrow })/2<h_{z}<U-U_{\uparrow \downarrow }$, and then
its size remains essentially unchanged for $h_{z}>U-U_{\uparrow \downarrow}$.
Similarly, when $h_z = 0$, Fig.~\ref{fig:t_0}(b) shows that $h_{y}$ has a
similar effect on the sizes of the MI lobes.
Having established the theoretical formalism, next we present the details of
our numerical calculations.

\section{Numerical Results}
\label{sec:nr}

First of all, we need to solve Eq.~(\ref{eqn:se})
self-consistently with Eqs.~(\ref{eqn:opup}) and~(\ref{eqn:opdo}) for the
eigenstates of the lowest-energy eigenvalue. This can be achieved via the
iterative method of relaxation as follows: first (i) start with an input set of $%
\lbrace \Delta _{j\alpha} \rbrace$, then (ii) construct the Hamiltonian matrix
given in Eq.~(\ref{eqn:se}), and then (iii) use the lowest-energy eigenstates in
Eqs.~(\ref{eqn:opup}) and~(\ref{eqn:opdo}) and generate a new set of $%
\lbrace \Delta _{j\alpha} \rbrace$, and finally (iv) repeat these steps
until the input and output sets of $\lbrace \Delta _{j\alpha} \rbrace$
lie within a confidence level. Once the iterative method converges,
we use Eqs.(\ref{eqn:ntot})-(\ref{eqn:py}) to calculate the local average
number of bosons and their polarizations.

As we emphasized in Sec.~\ref{sec:GA}, while the Gutzwiller ansatz
does not tell anything about the possibility of having magnetic spin
textures inside the MI lobes, it may still give a quantitatively accurate
description of the competition between the interaction, tunneling, SOC
and Zeeman field. In this section, we solve
Eqs.(\ref{eqn:ntot})-(\ref{eqn:se}) self-consistently near the MI-SF
phase transition boundary of the first two ($N = 1$ and 2) insulating lobes,
and analyze how their sizes change with the strength and
symmetry of the SOC and Zeeman field. For this purpose, we set the cut-off
of $l_\alpha$ in the sums to $l_\text{max} = 4$ which is sufficient
near the $N = 1$ and $N = 2$ MI lobes, and the intra-particle and
inter-particle interactions to $U_{\uparrow \uparrow}=U_{\downarrow
\downarrow}=U$ and $U_{\uparrow \downarrow }=0.3U$, respectively. In
addition, we choose equal magnitudes for the Rashba and ERD SOCs such that
$\gamma_E = \sqrt{2} \gamma_R$. It turns out that
Eqs.(\ref{eqn:ntot})-(\ref{eqn:se}) allow for many multiple solutions,
and therefore, we use of the order of 10$^{4}$ random initial sets
of $\lbrace \Delta_{j\alpha} \rbrace$, and then eventually keep the
one which has the lowest ground-state energy.

\subsection{SOC induced nonuniform SF phase}
\label{sec:SF}

In order to characterize the possible SF phases, we first solve the
self-consistency equations on finite $M \times L$ lattices with periodic
boundary conditions, but without any assumption on the symmetry of
$\Delta_{j\alpha }$. By letting $\lbrace M, L \rbrace= \{3, 4, 5, \cdots, 20\}$
and using numerous combinations of SOC and Zeeman fields, we find
that while the magnitudes of $\Delta _{j\alpha }$ are uniform across the
entire lattice, their phases may vary from site to site due to SOC, such that
\begin{equation}
\Delta _{j\alpha }=\overline{\Delta }_{\alpha }e^{i\theta _{j\alpha }},
\label{eqn:uop}
\end{equation}
where $\overline{\Delta }_{\alpha }=|\Delta _{j\alpha }|$ for all $j$. This
result is in agreement with an earlier study~\cite{mandal12}, and it shows
that the SF phase can be nonuniform depending on the model parameters.
Moreover, assuming Eq.~(\ref{eqn:uop}) holds, we solve the self-consistency
equations on very large lattices, and find that the phase $\theta _{j\alpha}$
jumps uniformly from one site to the next in $x$ and/or $y$ directions,
and also that the amount of jump is the same for both $\uparrow $ and $%
\downarrow $ components. In other words, equal-phase-jump configuration
between nearest-neighbor sites is energetically more favorable than the
repeating patterns of multiple phase jumps. Thus, our numerical calculations
suggest that the phases $\theta _{j\alpha }$, in their minimum-energy
configuration, obey the following pattern
\begin{equation}
\theta _{j\alpha }=\theta _{0\alpha }+j_{x}\theta _{x}+j_{y}\theta _{y}
\label{eqn:angle}
\end{equation}
where $(j_{x},j_{y})$ are $(x,y)$ coordinates of the site $j$ with respect
to our reference site $0$. In this paper, we set $\theta _{0\uparrow }=0$
without loosing generality, and determine the rest of the phases, i.e.,
$\theta _{0\downarrow }$, $\theta _{x}$ and $\theta _{y}$, with respect to it.
It also turns out that $\theta _{y}=0$ for ERD$_{x}$ and $\theta _{x}=0$
for ERD$_{y}$ SOC, and $|\theta _{x}|$ and $|\theta _{y}|$ are not
necessarily equal for Rashba SOC when $h_{y} \ne 0$.

\begin{figure}[h]
\begin{center}
\includegraphics[width=220pt]{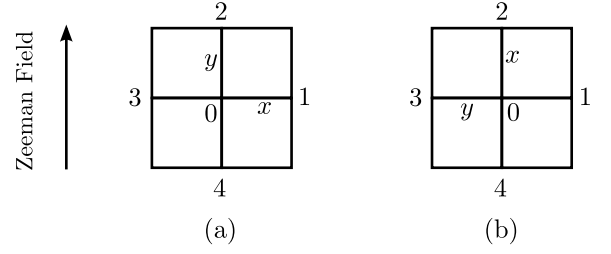}
\end{center}
\caption{
The phase $\theta _{0\downarrow }$ of the order parameter
$\Delta _{0\downarrow}$ of the reference site 0 may depend on the choice
of coordinate system, and is not a gauge-independent quantity within the
mean-field theory. This can be seen by comparing the order parameters using
the coordinate systems shown in (a) and (b) as discussed in the text.
}
\label{fig:L}
\end{figure}

Before we move on to the numerical analysis of the nonuniform-SF phases,
we emphasize that $\theta_{0\downarrow}$ may not be a gauge-independent
quantity due to the mean-field definition of the SF order parameters.
For illustration purposes, let us consider a lattice model with Rashba or
ERD SOC (the latter can be either parallel or perpendicular to the
in-plane Zeeman field), and write down its SF order parameters using
the two coordinate systems shown in Fig.~\ref{fig:L}.
The Hamiltonian of the system in (b) can be transformed to that of
the system in (a) via the following canonical transformation: $\hat{b}
_{j\uparrow}=\hat{a}_{j\uparrow }$ and $\hat{b}_{j\downarrow }=i\hat{a}
_{j\downarrow }$ for all $j$. Letting $|\psi ^{a}\rangle $ and $|\psi
^{b}\rangle $ be the ground states of (a) and (b), respectively, and
expanding $|\psi ^{a}\rangle$ in the occupation number basis $|l_{\uparrow
},l_{\downarrow }\rangle ^{a}$ of $a$-bosons and $|\psi ^{b}\rangle $ in
$|l_{\uparrow},l_{\downarrow }\rangle ^{b}$ of $b$-bosons,
show that the expansion coefficients are equal for the corresponding terms.
Therefore, the order parameter $\Delta_{0\downarrow
}^{a}$ of the reference site in (a) $\Delta_{0\downarrow }^{a}=\langle \psi
^{a}|a_{0\downarrow }|\psi ^{a}\rangle =\langle \psi ^{a}|b_{0\downarrow
}|\psi ^{a}\rangle $ is related to the order parameter of the same site in
(b) by $\Delta _{0\downarrow}^{b}=\langle \psi ^{b}|a_{0\downarrow }|\psi
^{b}\rangle =-i\langle \psi^{b}|b_{0\downarrow }|\psi ^{b}\rangle =-i\Delta
_{0\downarrow }^{a}$. This shows that $\theta _{0\downarrow}$ depends
on the coordinate system, and is not a gauge-independent quantity.
However, relative phases of all of the neighboring sites,
i.e., $\theta _{j\alpha}-\theta _{0\alpha}$ for all $j$, are not affected
by the above transformation, and hence, they are gauge independent.

Equations~(\ref{eqn:uop}) and~(\ref{eqn:angle}) suggest that our numerical
results (to be discussed below) for the phases $\theta_{0\downarrow}$,
$\theta _{x}$ and $\theta _{y}$ can be analytically understood by simply
looking at the local ground-state energy
$E_{0}=\langle \psi ^{j}|H|\psi ^{j}\rangle$ of the system at any particular site $j$.
Note that the local Gutzwiller
ground-state of site $j$, $|\psi ^{j}\rangle =\sum_{l_{\uparrow
}l_{\downarrow }}f_{l_{\uparrow }l_{\downarrow }}^{j}|l_{\uparrow
},l_{\downarrow }\rangle _{j}$, can be determined by solving Eq.~(\ref%
{eqn:se}) for the minimum-energy configuration. Similar to the SF order
parameters shown in Eq.~(\ref{eqn:uop}), our numerical calculations also
suggest that the magnitudes of $f_{l_{\uparrow }l_{\downarrow }}^{j}$ are
uniform across the entire lattice, such that
\begin{equation}
f_{l_{\uparrow }l_{\downarrow }}^{j}=\overline{f}_{l_{\uparrow
}l_{\downarrow }}e^{i\phi _{l_{\uparrow }l_{\downarrow }}^{j}},
\label{eqn:fp}
\end{equation}
where $\overline{f}_{l_{\uparrow }l_{\downarrow }}=|f_{l_{\uparrow
}l_{\downarrow }}^{j}|$ for all $j$. In addition, we find that while the
interaction terms compete with the the rest of the (tunneling, SOC and
in-plane Zeeman) terms in the Hamiltonian for the magnitudes
$\overline{f}_{l_{\uparrow}l_{\downarrow}}$,
the phases $\phi _{l_{\uparrow}l_{\downarrow}}^{j}$ are solely determined
by the interplay between tunneling, SOC and in-plane Zeeman field,
in such a way to minimize the energy $E_{0}$ for a
given set of magnitudes $\overline{f}_{l_{\uparrow }l_{\downarrow }}$. Using
Eqs.~(\ref{eqn:uop})-(\ref{eqn:fp}), and after some algebra,
$E_{0}=\langle \psi ^{0}|H|\psi^{0}\rangle$ of the reference site 0
can be written as
\begin{eqnarray}
E &&_{0}=-4t\left( \overline{\Delta }_{\uparrow }^{2}+\overline{\Delta }%
_{\downarrow }^{2}\right) (\cos \theta _{x}+\cos \theta _{y})  \notag \\
&&-8\overline{\Delta }_{\uparrow }\overline{\Delta }_{\downarrow }\left(
\gamma _{y}\cos \theta _{0\downarrow }\sin \theta _{y}-\gamma _{x}\sin
\theta _{0\downarrow }\sin \theta _{x}\right)  \notag \\
&&+\sum_{l_{\uparrow }l_{\downarrow }}\overline{f}_{l_{\uparrow
}l_{\downarrow }}^{2}\left\{ U_{\uparrow \downarrow }l_{\uparrow
}l_{\downarrow }+ \sum_{\alpha } \left[ \frac{U_{\alpha \alpha }}{2}l_{\alpha
}(l_{\alpha }-1)-\mu _{\alpha }l_{\alpha }\right] \right\}  \notag \\
&&-2h_{y}\text{Im}\sum_{l_{\uparrow }l_{\downarrow }} f_{l_{\uparrow
}l_{\downarrow }}^{0\ast } f_{l_{\uparrow }-1,l_{\downarrow }+1}^{0} \sqrt{%
l_{\uparrow }(l_{\downarrow }+1)}.  \label{eqn:0energy}
\end{eqnarray}
Using Eq.~(\ref{eqn:py}), the last term can also be written as $-h_y P_{0y}$.
Much of our analytical understanding of the numerical calculations
is based on the analysis of this expression in various limits, and we
refer to it quite frequently in the remaining parts of the paper.

For example, in the simpler case of non-SF phases when $\overline{\Delta }%
_{\alpha} = 0$, since the tunneling and SOC terms disappear
from Eq.~(\ref{eqn:0energy}), the set of phases
$\lbrace \phi_{l_{\uparrow }l_{\downarrow }}^{j} \rbrace$
is determined only by $h_y$. For a given set of $\{\overline{f}
_{l_{\uparrow}l_{\downarrow }}\}$, the contribution of the in-plane
Zeeman field to $E_{0}$ is minimum when the relative angles satisfy the
condition $\phi _{l_{\uparrow}-1,l_{\downarrow }+1}^{0} - \phi _{l_{\uparrow
}l_{\downarrow }}^{0}=\pi/2$ for all $l_{\uparrow }$ and $l_{\downarrow}$ as
long as $h_y \ne 0$.

This condition still holds in the SF phase as long as $h_y \ne 0$ and there is
no SOC. To prove this, let us set $\gamma_x = \gamma_y = 0$ in
Eq.~(\ref{eqn:0energy}), in which case contribution of the tunneling term
$-4t(\overline{\Delta }_{\uparrow }^{2}+ \overline{\Delta }_{\downarrow }^{2})
(\cos \theta_{x}+\cos \theta _{y})$ to $E_0$ is minimized when $\overline{\Delta }
_{\uparrow }$ and $\overline{ \Delta }_{\downarrow }$ are maximum. This is
because $\theta _{x}=\theta _{y}=0$ when there is no SOC, leading to
a uniform SF phase across the entire lattice. Using Eq.~(\ref{eqn:opup}),
we have
$\Delta _{0\uparrow } =\sum_{l_{\uparrow }l_{\downarrow }}\overline{f}
_{l_{\uparrow }l_{\downarrow }}\overline{f}_{l_{\uparrow }+1,l_{\downarrow
}}e^{i(\phi _{l_{\uparrow }+1,l_{\downarrow }}^{0}-\phi _{l_{\uparrow
}l_{\downarrow }}^{0})}\sqrt{l_{\uparrow }+1} $ which is chosen to be
a real number in this paper, but $\Delta _{0\downarrow } =\sum_{l_{\uparrow
}l_{\downarrow }}\overline{f} _{l_{\uparrow }l_{\downarrow }}\overline{f}%
_{l_{\uparrow },l_{\downarrow }+1}e^{i(\phi _{l_{\uparrow },l_{\downarrow
}+1}^{0}-\phi _{l_{\uparrow }l_{\downarrow }}^{0})}\sqrt{l_{\downarrow }+1} $
is a complex number in general.
Therefore, for a given set of $\{\overline{f}_{l_{\uparrow }l_{\downarrow
}}\}$, the order parameters are maximized when $\phi _{l_{\uparrow
}+1,l_{\downarrow }}^{0} -\phi_{l_{\uparrow }l_{\downarrow }}^{0}=const$ and
$\phi _{l_{\uparrow },l_{\downarrow }+1}^{0}-\phi _{l_{\uparrow
}l_{\downarrow }}^{0}=const$ for all $l_{\uparrow }$ and $l_{\downarrow}$.
Recall that since we already set $\theta _{0\uparrow }=0$ in this paper, $%
\phi_{l_{\uparrow }+1,l_{\downarrow }}^{0}-\phi _{l_{\uparrow }l_{\downarrow
}}^{0}=0$ maximizes the order parameters. It is important to note that a
set of phases can simultaneously satisfy both this condition and the
condition $\phi _{l_{\uparrow }-1,l_{\downarrow }+1}^{0}-\phi _{l_{\uparrow
}l_{\downarrow }}^{0}$ $=\pi /2$ that minimizes the in-plane Zeeman term,
and combining these two conditions reveals that $\phi_{l_{\uparrow},l_{%
\downarrow }+1}^{0} -\phi _{l_{\uparrow }l_{\downarrow }}^{0}=\pi /2$ for
all $l_{\uparrow }$ and $l_{\downarrow }$. This in turn implies that
$\theta_{0\downarrow }=\pi /2$, which is in agreement with our
numerical results.

For completeness, here we find it is useful to comment on the effects of
an in-plane $h_x$ Zeeman field. If such a field is considered in
Eq.~(\ref{eqn:ham}), its contribution to $E_0$ can be explicitly written
as $-2h_{x} \text{Re} \sum_{l_{\uparrow }l_{\downarrow
}} f_{l_{\uparrow }l_{\downarrow }}^{0\ast} f_{l_{\uparrow }-1,l_{\downarrow
}+1}^{0}\sqrt{l_{\uparrow }(l_{\downarrow}+1)}$, which
is nothing but $-h_x P_{0x}$. Assuming $h_y = 0$, and using similar arguments
as above, we find $\theta _{0\downarrow}=0$ in this case, which is again
in agreement with our numerical results. Note that this analysis
is also consistent with our previous discussion about the
relation between the order parameters $\Delta_{0\downarrow}^{b}=-i\Delta
_{0\downarrow }^{a}$ that are defined using the coordinate systems shown in
Fig.~\ref{fig:L}. Next, we are ready to analyze the effects of the strength and
symmetry of the SOC and Zeeman field on the nonuniform-SF phase
and resultant phase diagrams.

\begin{figure*}[t]
\begin{center}
\includegraphics[width=450pt]{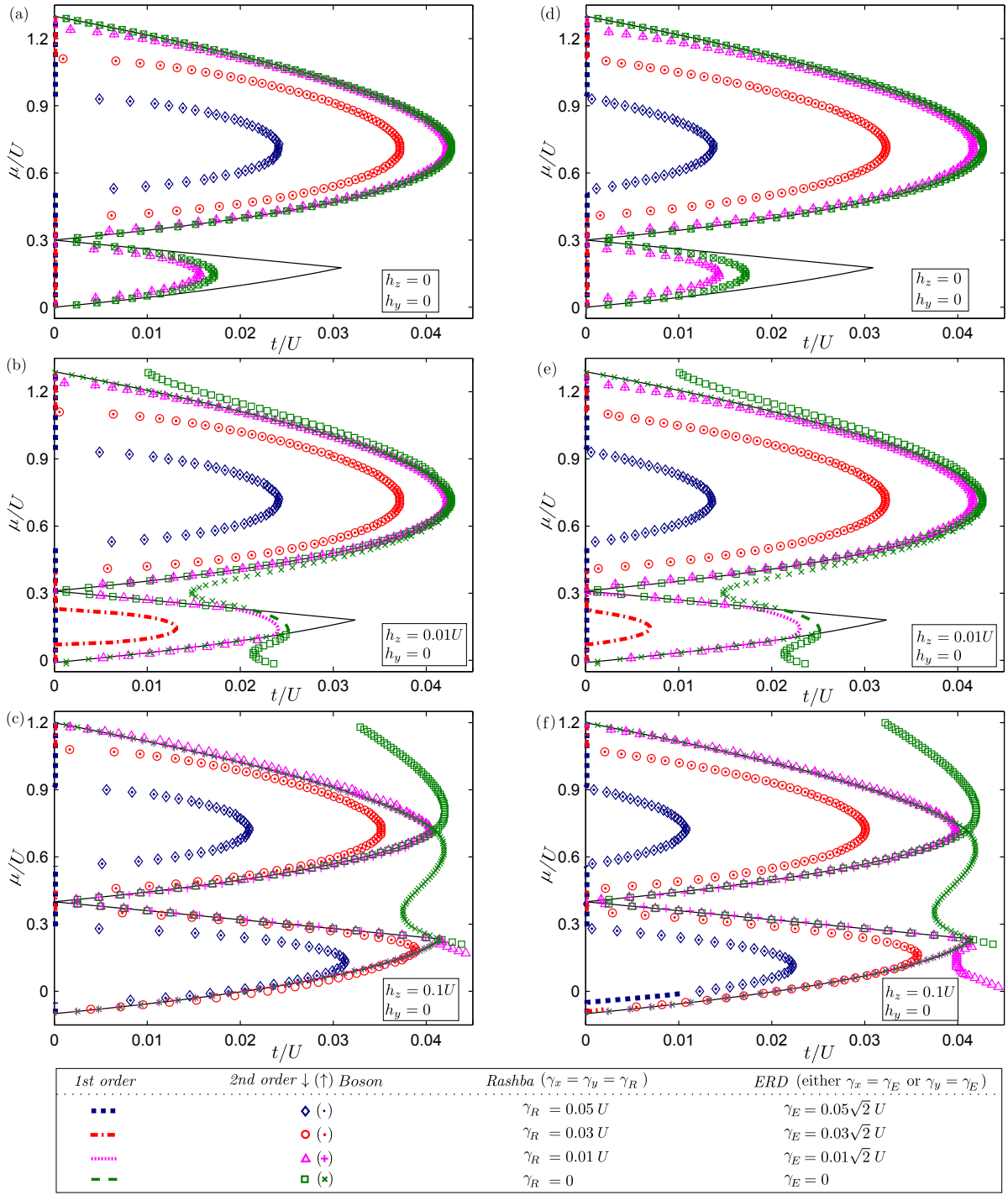}
\end{center}
\caption{(Color online)
Ground-state phase diagrams with out-of-plane
$h_z$ Zeeman fields. The MI-SF phase transition boundaries are shown as functions
of $\mu$ and $t$ for the first two MI lobes, i.e., $N=1$ and 2,
where we consider Rashba SOC in (a)-(c), and ERD SOC in (d)-(f). Here, we
set $h_y = 0$, $U_{\uparrow \uparrow}=U_{\downarrow \downarrow}=U$ and $%
U_{\uparrow \downarrow}=0.3U$ in all figures. In addition, the black solid
lines are guides to the eye, which are obtained from Eq.~(\ref{eqn:black})
(see the text for details).
}
\label{fig:1}
\end{figure*}
\subsection{MI-SF phase transitions: out-of-plane Zeeman field}
\label{sec:outZ}

We set $h_y = 0$ in this section, and study Rashba and ERD SOCs with an
out-of-plane Zeeman field. We recall that the SF phases in this work are
distinguished from the MI ones by their finite $\overline{\Delta}_\uparrow$
and/or $\overline{\Delta}_\downarrow$ order parameters, and therefore,
single-particle/single-hole excitations are always gapped inside the MI
lobes. However, since our definition of the SF order parameters does not
discriminate the possibility of exotic multi particle and/or hole
excitations that may be gapless, our single-particle/single-hole MI lobes
may still have some sort of hidden (exotic) SF orders. The fate of such
exotic SF phases is beyond the scope of this work, and they
deserve a separate analysis on their own (see also Sec.~\ref{sec:conc}).

In Figs.~\ref{fig:1}(a) and~(d), we show the $h_{z}=0$ ground-state
phase diagrams as functions of $\mu $ and $t$ for Rashba and ERD
SOCs, respectively.
Since $U_{\uparrow \uparrow }=U_{\downarrow \downarrow }$ and $\mu
_{\uparrow }=\mu _{\downarrow }$ in these figures, the order parameters
must also be equal $\overline{\Delta }_{\uparrow }=\overline{\Delta }_{\downarrow}$,
and therefore, both $\alpha $ components simultaneously undergo MI-SF
transition across the phase transition boundary. In particular, the $N=2$ MI
lobe is characterized by $N_{\uparrow }=N_{\downarrow }=1$ and all of its
elementary excitations are gapped. However, the $N=1$ lobe is proposed to
have an exotic counterflow-SF order of particle-hole pairs as discussed in
the literature when there is no
SOC~\cite{kuklov03, altman03, hu09, ozaki12}.
These figures clearly show that the sizes of MI lobes shrink as
a function of increasing SOC strength in both Rashba and ERD cases, which is
a result of increased mobility of the particles due to SOC tunneling.
Note in Figs.~\ref{fig:1}(a) and~(d) that the $N=1$ MI lobe shrinks
so much that it lives right on the $\mu $-axis for sufficiently strong SOC,
and the system becomes a SF even in the $t/U\rightarrow 0$ limit.

In Fig.~\ref{fig:1} (and also the ones below), the black solid lines are guides
to the eye, and they represent the MI-SF phase transition boundary between
the $(N_\uparrow, N_\downarrow)$ MI lobes and uniform-SF phase when there is
no SOC and in-plane Zeeman field. Setting $\gamma_x = \gamma_y = 0$
and $h_y = 0$ in Eq.~(\ref{eqn:ham}), the mean-field MI-SF phase transition
boundary can easily be obtained within the decoupling approximation, leading
to the analytical expression~\cite{iskin10},
\begin{eqnarray}
\mu _{\alpha }^{p,h} &=&U_{\alpha \alpha}(N_{\alpha }-1/2)+U_{\uparrow
\downarrow }N_{-\alpha }-2t  \notag \\
&&\pm \sqrt{U_{\alpha \alpha}^{2}/4-U_{\alpha \alpha}(4N_{\alpha }+2)t+4t^{2}},
\label{eqn:black}
\end{eqnarray}
where $\alpha \equiv (\uparrow, \downarrow)$ labels the transition from $%
\alpha$-MI to $\alpha$-SF, and $(p,h)$ together with $\pm$ signs correspond
to the particle and hole branches, respectively. Here, $(-\uparrow) \equiv
\downarrow$ and vice versa. Note when $U_{\uparrow \downarrow} = 0$
that Eq.~(\ref{eqn:black}) reduces to two independent copies of the usual
mean-field result for the single-component model. We use
$\mu_{\uparrow}^{h}$ for the transition from $(1,0)$ MI to $\uparrow$%
-SF by removing one $\uparrow$ boson, $\mu _{\downarrow }^{p} $
for the transition from $(1,0)$ MI to $\downarrow$-SF by adding one $%
\downarrow$ boson, $\mu _{\downarrow }^{h}$ for the transition
from $(1,1)$ MI to $\downarrow$-SF by removing one $\downarrow$ boson, $%
\mu_{\uparrow }^{p}$ for the transition from $(1,1)$ MI to $%
\uparrow$-SF by adding one $\uparrow$ boson, etc. However, this
expression is not applicable to the $N = 1$ MI lobe when there is no
SOC and Zeeman field, and this is clearly seen in Fig.~\ref{fig:1}(a) and~(d).

In Figs.~\ref{fig:1}(b) and~(e), a relatively small $h_z = 0.01U$ is added
to the parameters of Figs.~\ref{fig:1}(a) and~(d), breaking the
degeneracy between $\uparrow$ and $\downarrow$ bosons.
Although this leads to a slight shift in $\mu_\uparrow$ and
$\mu_\downarrow$, it has dramatic consequences on the ground-state
phase diagrams. First of all, unlike the $h_z = 0$ case,
the $\uparrow$ and $\downarrow$ bosons do not simultaneously
become SF, unless the spin-mixing SOC strength is sufficiently
strong and makes up for the chemical potential asymmetry caused by
$h_z \ne 0$. For instance, near the particle (hole) branch of the
$N = 1$ MI lobe, it is the $\downarrow$- ($\uparrow$)-component
which first becomes a $\downarrow$-SF ($\uparrow$-SF) as a
function of increasing $t$.
Second, unlike the $h_z = 0$ case, the $N = 1$ MI lobe
becomes a $(1,0)$ MI, and therefore, its SF transition boundary
gradually converges to that given by Eq.~(\ref{eqn:black}) for
small $t/U$ values.
Third, unlike the $h_z = 0$ case, we find regions of first-order
MI-SF phase transitions near the tip of the $N = 1$ lobe, and this
explains why the boundary given by our numerical calculations and
Eq.~(\ref{eqn:black}) have an increasing mismatch for large
$t/U$ values.

When $h_z$ is increased to $h_z = 0.1U$, as shown in Figs.~\ref{fig:1}(c)
and~(f), the first-order transition regions shrink near to the very tip
of the $N = 1$ MI lobe, and therefore, Eq.~(\ref{eqn:black}) provides
better matches with our numerical results. We also note that while
the size of the $N = 1$ MI lobe grows with increasing $h_{z}$, the size of
the $N = 2$ MI lobe shrinks. We find that such trends are independent of
SOC strength, and therefore, they are in good qualitative
agreement with what is expected from Eq.~(\ref{eqn:black}). Furthermore,
since these trends are also strongly correlated with the sizes of the MI
lobes in the atomic limit, Fig.~\ref{fig:t_0} provides a rough intuition
about how the sizes of the MI lobes change as a function of $h_{z}$.

In addition, we see in Fig.~\ref{fig:1} that ERD SOC shrinks the sizes of MI
lobes a little bit more than Rashba SOC, and this small difference may be
understood from Eq.~(\ref{eqn:0energy}) as follows. For a given set of $\{%
\overline{f}_{l_{\uparrow }l_{\downarrow }}\}$, contribution of the
tunneling term to $E_{0}$ is minimized when $\overline{\Delta }_{\uparrow }$
and $\overline{\Delta }_{\downarrow }$ are maximum, and this can at best be
achieved if $\phi _{l_{\uparrow }+1,l_{\downarrow }}^{0}-\phi _{l_{\uparrow
}l_{\downarrow }}^{0}=const=\theta _{0\uparrow }$ and $\phi _{l_{\uparrow
},l_{\downarrow }+1}^{0}-\phi _{l_{\uparrow }l_{\downarrow
}}^{0}=const=\theta _{0\downarrow }$ for all $l_{\uparrow }$ and $%
l_{\downarrow }$. Recall that the former phase is set to 0 in this paper
[see the discussion below Eq.~(\ref{eqn:angle})]. 
Minimizing the contribution of Rashba SOC terms to $E_{0}$, 
we find four-fold degenerate solutions:
\begin{tabular}[c]{|c|c|c|c|c|}
\hline
 & (i) & (ii) & (iii) & (iv) \\
\hline
$\ \ \theta _{0\downarrow }=\ \ $ & $\ \ \pi /4\ \ $ & $\ \ 3\pi /4\ \ $ & $\ \ -3\pi /4\ \ $ & $\ \ -\pi /4\ \ $ \\
%\hline
$\theta _{x}\in $ & $\ \ [-\pi /2,0)\ \ $ & $\ \ [-\pi /2,0)\ \ $ & $\ \ \ (0,\pi /2]\ \ \ $ & $\ \ (0,\pi
/2]\ \ $ \\ %\hline
$\theta _{y}\in $ & $(0,\pi /2]$ & $[-\pi /2,0)$ & $[-\pi /2,0)$ & $(0,\pi
/2]$ \\ \hline
\end{tabular}
where the semi-open intervals are due to non-zero SOC. On the other hand, since
the ERD SOC breaks the rotational symmetry, we set $\theta _{x}=\theta _{E}$
and $\theta _{y}=0$ for ERD$_{x}$, and $\theta _{y}=\theta _{E}$ and $\theta
_{x}=0$ for ERD$_{y}$ SOC. Minimizing the contribution of ERD$_{x}$ SOC term
to $E_{0}$, we find two-fold degenerate solutions: (i) $\theta _{E}\in
(0,\pi /2]$ together with $\theta _{0\downarrow }=\pi /2$, and (ii) $\theta _{E}\in
\lbrack -\pi /2,0)$ together with $\theta _{0\downarrow }=-\pi /2$. Similarly,
minimizing the contribution of ERD$_{y}$ SOC term to $E_{0}$, we again find
two-fold degenerate solutions: (i) $\theta _{E}\in (0,\pi /2]$ together with $\theta
_{0\downarrow }=0$, and (ii) $\theta _{E}\in \lbrack -\pi /2,0)$ together with
$\theta _{0\downarrow }=\pi $. Based on this analysis, the tunneling and
SOC contributions to $E_0$ can be written as,
$-8t(\overline{\Delta }_{\uparrow }^{2}+\overline{
\Delta }_{\downarrow }^{2})\cos \theta _{R}-8\overline{\Delta }_{\uparrow }
\overline{\Delta }_{\downarrow }\gamma _{R}\sqrt{2}\sin \theta _{R}$ for
the Rashba SOC where $|\theta_{x}|=|\theta _{y}|=\theta _{R}\in(0,\pi/2]$, and $-4t(\overline{\Delta }_{\uparrow }^{2}+\overline{\Delta
}_{\downarrow }^{2})(1+\cos \theta _{E}) -8\overline{\Delta }_{\uparrow }
\overline{\Delta }_{\downarrow }\gamma _{E}\sin |\theta _{E}|$ for the ERD
SOC. Setting $\gamma _{E}=\sqrt{2}\gamma_{R}$ as in our numerical
calculations, these expressions show that the contribution of ERD SOC to
$E_0 $ is always smaller than that of Rashba SOC, which in turn implies that
ERD SOC gives way to SF phase for smaller $t$ values.

Before we move on to the next section, we remark that minimizing these
contributions with respect to $\theta_R$ and $\theta_E$, we obtain
$
\tan \theta _{R}=\frac{\sqrt{2}\gamma _{R}
\overline{\Delta }_{\uparrow }\overline{\Delta }_{\downarrow }}{t(\overline{
\Delta }_{\uparrow }^{2}+\overline{\Delta }_{\downarrow }^{2})}
$
for the Rashba SOC, and $\tan |\theta _{E}|=\frac{2\gamma _{E}\overline{\Delta }
_{\uparrow } \overline{\Delta }_{\downarrow }}{t(\overline{\Delta }
_{\uparrow }^{2}+ \overline{\Delta }_{\downarrow }^{2})}
$
for the ERD SOC, respectively. In the simplest case when $h_{z}=0$,
setting $\overline{\Delta}_{\uparrow }=\overline{\Delta }_{\downarrow }$
leads to $\tan \theta_{R}=\gamma _{R}/(t\sqrt{2})$ for the Rashba SOC
and $\tan|\theta_{E}|=\gamma _{E}/t$ for the ERD SOC, which are in good
agreement with our numerical results. In particular, we checked that
$\tan |\theta _{E}|=2\tan \theta_{R}$ is satisfied in Figs.~\ref{fig:1}(a) and~(d)
for any given $\mu $ and $t$ as long as both ground-states are SF.
Next, we are ready to analyze the ground-state phase diagrams
in the presence of a general Zeeman field.

\begin{figure*}[t]
\begin{center}
\includegraphics[width=450pt]{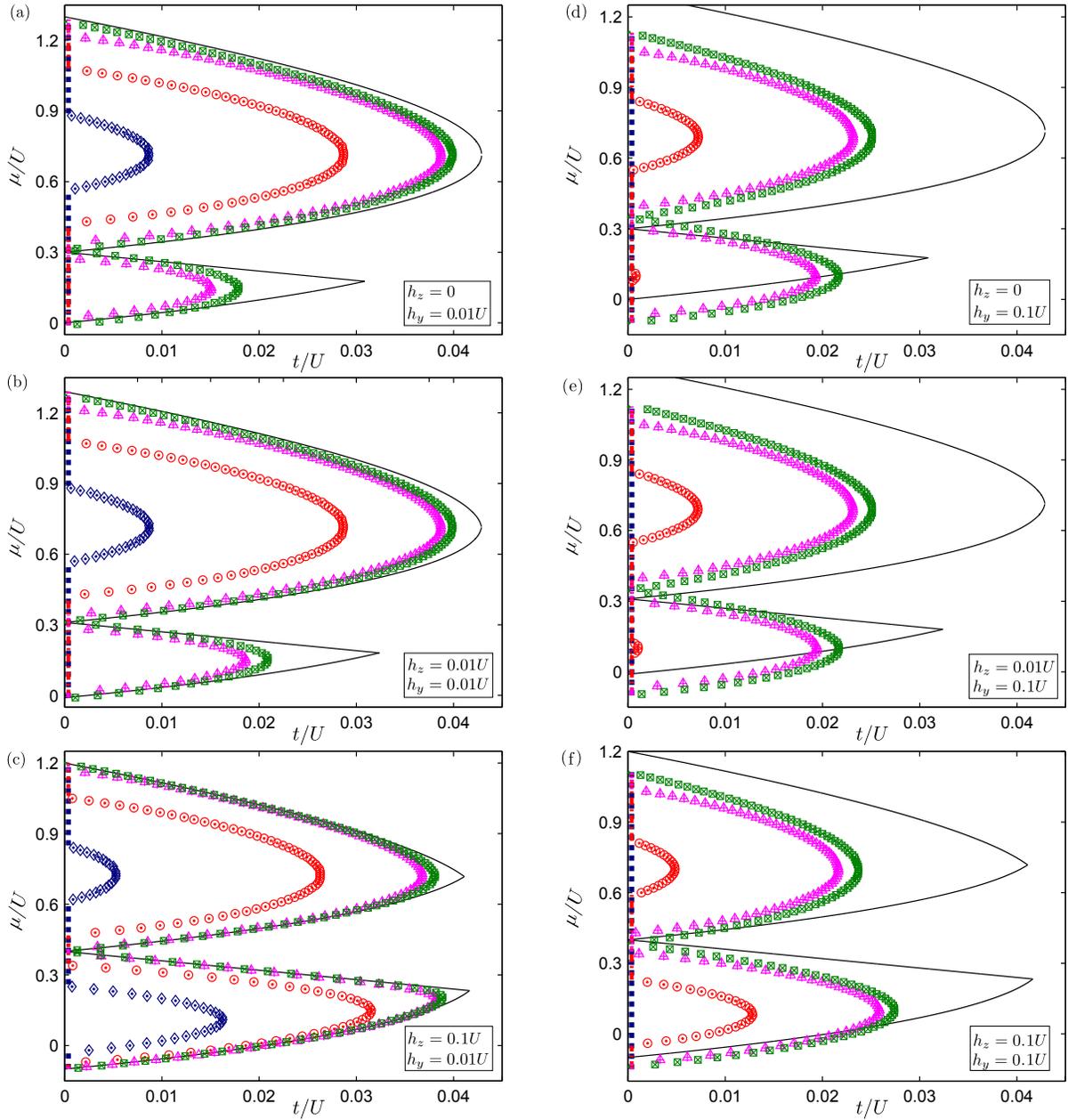}
\end{center}
\caption{(Color online)
Ground-state phase diagrams with ERD$_x$ SOCs ($
\gamma _{x}=\gamma _{E}, \gamma _{y}=0$) and general
Zeeman fields. Rest of the parameters are specified in Fig.~\ref{fig:1}.
}
\label{fig:2}
\end{figure*}
\begin{figure*}[t]
\begin{center}
\includegraphics[width=450pt]{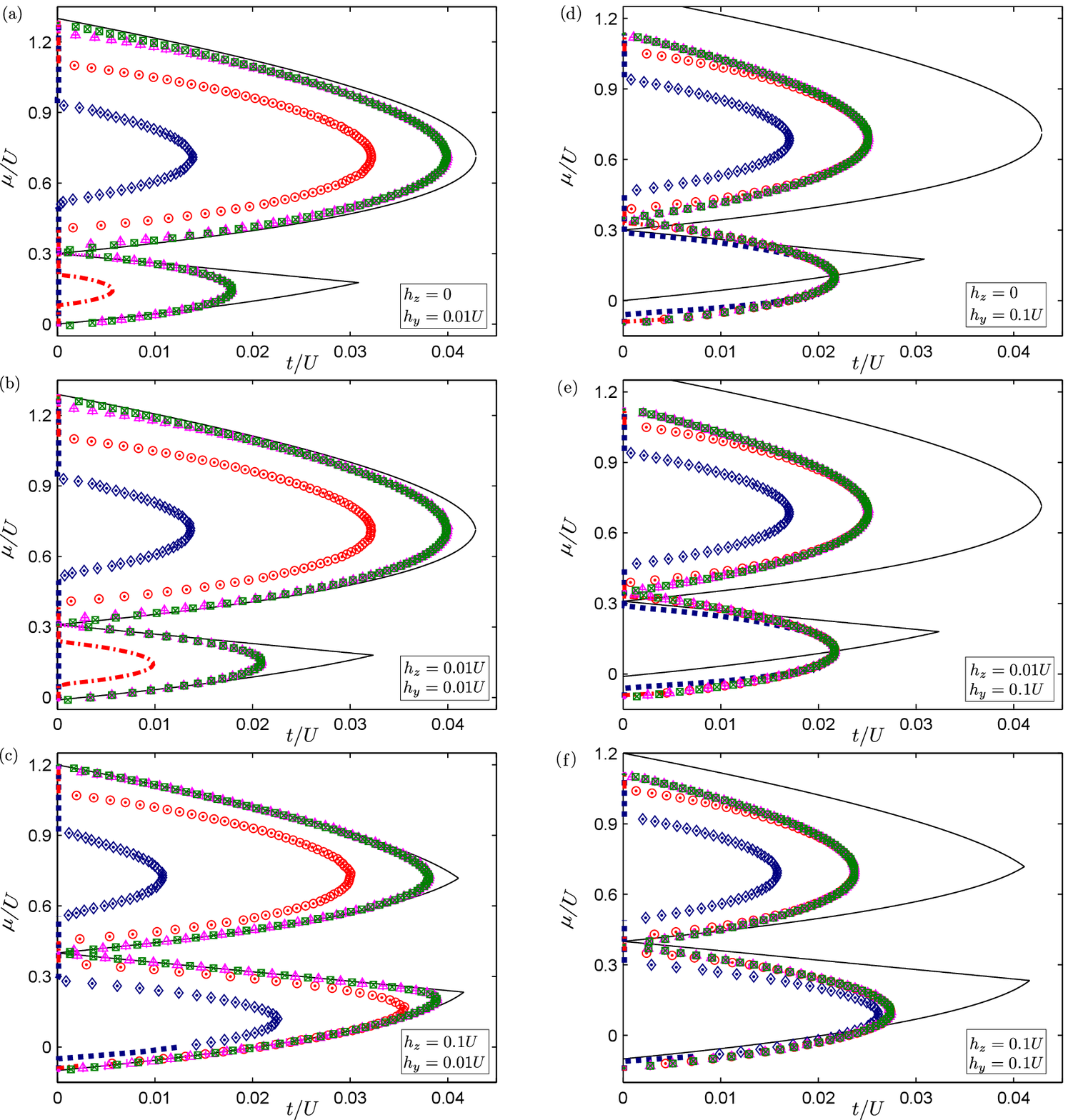}
\end{center}
\caption{(Color online)
Ground-state phase diagrams with ERD$_y$ SOCs
($\gamma _{x}=0, \gamma _{y}=\gamma _{E}$) and general
Zeeman fields. Rest of the parameters are specified in Fig.~\ref{fig:1}.
}
\label{fig:3}
\end{figure*}
\begin{figure*}[t]
\begin{center}
\includegraphics[width=450pt]{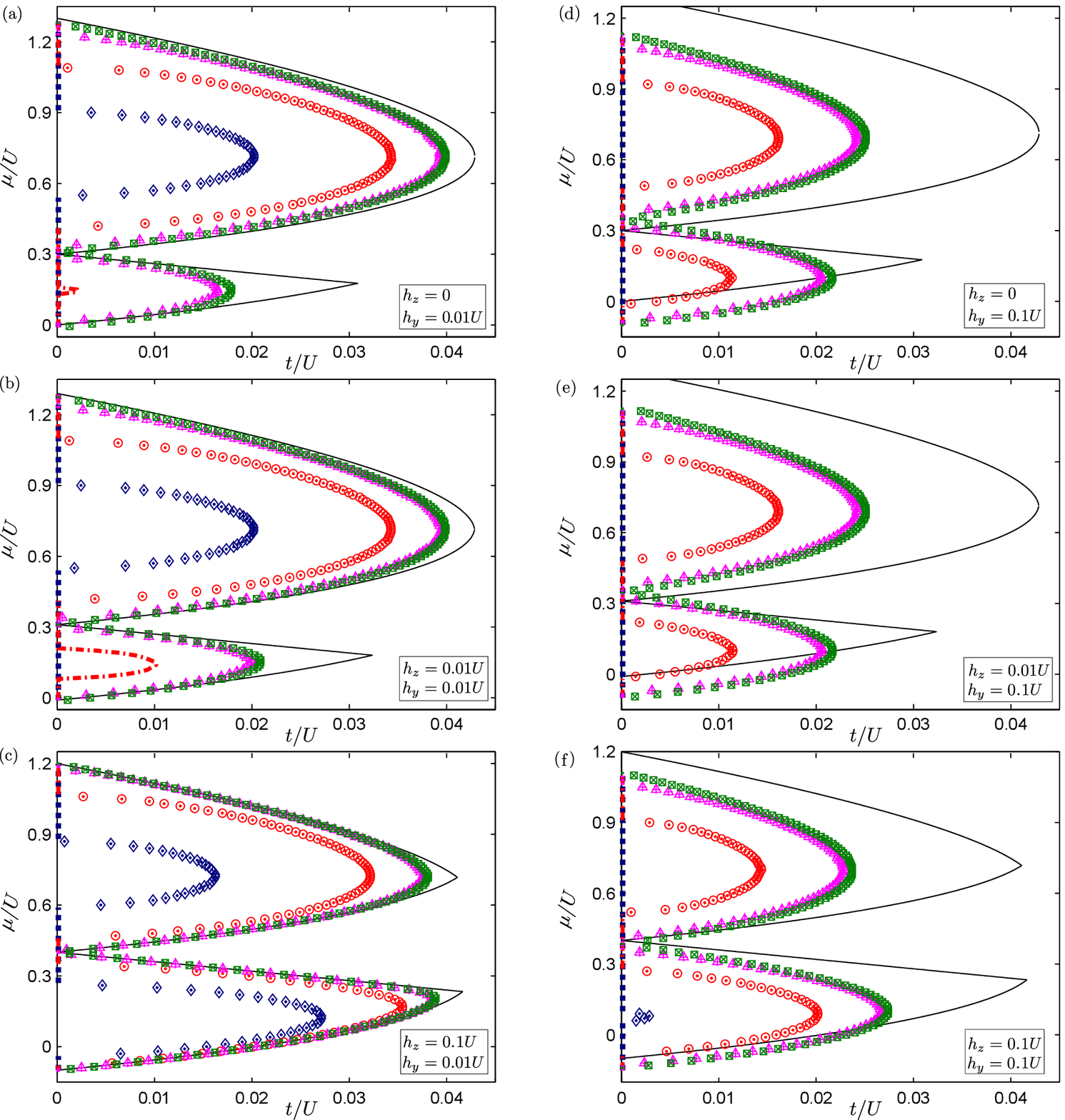}
\end{center}
\caption{(Color online)
Ground-state phase diagrams with Rashba SOCs
($\gamma _{x}=\gamma _{y}=\gamma _{R}$) and general
Zeeman fields. Rest of the parameters are specified in Fig.~\ref{fig:1}.
}
\label{fig:4}
\end{figure*}
\subsection{MI-SF phase transitions: general Zeeman field}
\label{sec:inZ}

So far, we argued when $h_y = 0$ that while the MI-SF phase transition
boundaries are essentially identical for ERD$_x$ and ERD$_y$ SOCs
(since they are related through a spin-rotation symmetry), their
nonuniform-SF phases may be characterized by gauge-dependent order
parameters. However, $h_y \ne 0$ breaks the symmetry between
ERD$_x$ and ERD$_y$ SOCs, and therefore, it is expected that
$h_y \ne 0$ has dramatic consequences on the ground-state phase
diagrams.

In Figs.~\ref{fig:2}-\ref{fig:4}, we show the $h_{y} \ne 0$ phase diagrams
as functions of $\mu $ and $t$ for ERD$_x$, ERD$_y$ and Rashba SOCs,
respectively. To understand the differences and similarities between these
diagrams, first of all, we recall when $h_y = 0$ and $h_z \ne 0$ that the
symmetry between $\uparrow$ and $\downarrow$ bosons is broken, and
therefore, they do not simultaneously become SF, unless the SOC strength
is sufficiently strong. In a somewhat similar fashion, we find
that increasing $h_y$ from 0 eventually causes simultaneous transition of
$\uparrow$ and $\downarrow$ bosons to SF, even with a relatively
small $h_{y}=0.01U$ for our chosen model parameters.
This is because both SOC and in-plane Zeeman field do not conserve
spin, and they couple $\uparrow$ and $\downarrow$ bosons.
However, the effects of on-site spin mixing due to $h_y$ is much
more stronger than that of the off-site spin mixing due to SOC.
In addition, we note in all Figs.~\ref{fig:2}-\ref{fig:4}
that the sizes of MI lobes shrink as a function of increasing SOC strength,
which is a result of increased mobility of $\alpha$ particles as mentioned
in the previous section. However, the relative sizes of the MI lobes
vary significantly depending on the symmetry of the SOC, and these
differences can be understood from Eq.~(\ref{eqn:0energy}) as follows.

In Sec.~\ref{sec:SF}, we argued when there is no SOC that $\theta
_{0\downarrow} = \pi /2$ as long as $h_y \ne 0$.
Next, we show that an ERD$_x$ SOC is not in competition with
$h_y$ for the value of $\theta _{0\downarrow}$, and this relation
still holds. As noted in the previous section, $\theta _{y} = 0$ for
ERD$_x$ SOC, and we find that $\theta _{0\downarrow }=\pi /2$ together
with $\theta _{x} \in [-\pi /2,0)$ minimize contribution of the
ERD$_x$ SOC to $E_0$. Note that this condition is in no conflict
with the tunneling term since $\cos(x)$ is an even function.
Minimizing the combined contributions of the tunneling and ERD$_x$
SOC, $-4t(\overline{\Delta }_{\uparrow }^{2}+\overline{\Delta }_{\downarrow
}^{2})(\cos \theta _{x}+1)+8\overline{\Delta }_{\uparrow }\overline{\Delta }
_{\downarrow }\gamma _{x}\sin \theta _{x}, $ with respect to $\theta _{x}$,
we obtain $\tan \theta _{x}=-\frac{2\gamma _{x}\overline{\Delta }_{\uparrow}%
\overline{\Delta } _{\downarrow }}{t(\overline{{\Delta }}_{\uparrow }^{2}+%
\overline{\Delta } _{\downarrow }^{2})}, $ which is similar to the
expression we find in Sec.~\ref{sec:outZ}. In the simplest case when $h_z = 0$%
, setting $\overline{\Delta }_{\uparrow} = \overline{\Delta }_{\downarrow}$
leads to $\tan \theta _{x} = -\gamma _{x}/t$. Thus, in the case of ERD$_{x}$
SOC, $h_y \ne 0$ clearly lifts the two-fold degeneracy of $\theta
_{0\downarrow }=\pm \pi /2$ solutions discussed in Sec.~\ref{sec:outZ}.

On the other hand, ERD$_{y}$ SOC is in competition with $h_y$
for the value of $\theta _{0\downarrow}$, which can be easily
inferred by looking at the two extreme limits. When $h_{y} \gg \gamma _{y}
\to 0$, we already show in Sec.~\ref{sec:SF} that $\theta _{0\downarrow} =
\pi/2$. However, when $\gamma _{y} \gg h_{y} \to 0$, we show in Sec.~\ref%
{sec:outZ} that the ground state is two-fold degenerate: both $\theta
_{0\downarrow} = 0$ and $\pi$ minimize $E_0$. Thus, $\theta _{0\downarrow}$
clearly depends on the ratio of $\gamma_y$ and $h_y$, and we may
write $\theta_{0\downarrow} = \pi/2 \pm \eta$, where $\eta$ is determined
by $\gamma_y/h_y$.
Setting $\theta _{x} =
0$ for the ERD$_{y}$ SOC, and minimizing the combined
contribution of tunneling and ERD$_{y}$ SOC terms, $-4t(\overline{ \Delta }%
_{\uparrow }^{2}+\overline{\Delta }_{\downarrow }^{2})(1+\cos \theta _{y})-8%
\overline{\Delta }_{\uparrow }\overline{\Delta }_{\downarrow }\gamma
_{y}\cos \theta _{0\downarrow }\sin |\theta _{y}|$, with respect to
$\theta_y$, we obtain $\tan|\theta _{y}|=\frac{2\gamma _{y}\cos \theta
_{0\downarrow }\overline{\Delta } _{\uparrow }\overline{\Delta }_{\downarrow
}}{t(\overline{\Delta }_{\uparrow }^{2}+\overline{\Delta }_{\downarrow }^{2})%
}. $ In the simplest case when $h_z = 0$, setting $\overline{\Delta }%
_{\uparrow} = \overline{\Delta }_{\downarrow}$ leads to $\tan |\theta
_{y}|=\gamma _{y}\cos \theta _{0\downarrow }/t$.
In sharp contrast to the ERD$_x$ SOC phase diagrams where a relatively
small $h_y=0.01U$ has sizable effects on the MI lobes as shown in
Figs.~\ref{fig:2}(a)-(c), it has negligible effect on the ERD$_y$ SOC
diagrams that are shown in Fig.~\ref{fig:3}(a)-(c). However, when the
Zeeman field is sufficiently strong such as $h_y=0.1U$, we see
in Fig.~\ref{fig:3} that ERD$_y$ SOC has negligible effects.

These findings may explain why the effects of $h_y \ne 0$ on the ground-state
phase diagrams of Rashba SOC are stronger (weaker) than those of
ERD$_{y}$ (ERD$_{x}$) SOC. This is because while the $h_y$ term
competes with the $\gamma _{y}$ component of the Rashba SOC for
the value of $\theta _{0\downarrow}$, it does not compete with the $\gamma _{x}$
component. Therefore, $h_y \ne 0$ has an intermediate effect on the phase
diagrams of the Rashba SOC. Similar to the ERD$_y$ SOC, it should not be
surprising that the Rashba SOC is also competing with the $h_y$ term for the
value of $\theta _{0\downarrow}$, which can again be easily inferred by
looking at the two extreme limits. When $h_{y} \gg \gamma _{R} \to 0$, we
already show in Sec.~\ref{sec:SF} that $\theta _{0\downarrow} = \pi/2$.
However, when $\gamma _{R} \gg h_{y} \to 0$, we show in Sec.~\ref{sec:outZ}
that the ground state is four-fold degenerate: both $\theta _{0\downarrow}
= \pm\pi/4$ and $\pm3\pi/4$ minimize $E_0$. Thus, $\theta _{0\downarrow}$ clearly
depends on the ratio of $\gamma_R$ and $h_y$, and we may write 
$\theta_{0\downarrow} = \pi/2 \pm \eta$, where $\eta \in (0,\pi/4)$ is determined
by $\gamma_R/h_y$. This discussion shows thet $h_y \ne 0$ reduces the 
four-fold degeneracy of $\theta _{0\downarrow}$ to two-fold. In addition, we also
note that $h_y \ne 0$ breaks the rotational $xy$-symmetry, and therefore,
$|\theta _{x}|$ and $|\theta _{y}|$ are not necessarily equal to each other.
This completes our analysis of in-plane Zeeman field on the ground-state
phase diagrams, and we are ready to conclude the paper with a brief
summary of our results and an outlook.

\section{Summary and Outlook}
\label{sec:conc}

To conclude, here we considered a square lattice in two dimensions
and studied the effects of both the strength and symmetry of SOC and
Zeeman field on the ground-state phases and phase diagram of the
two-component Bose-Hubbard model. In particular, based on a
variational Gutzwiller ansatz, we analyzed the competition between the
interaction, tunneling, Rashba and ERD SOCs, and out-of- and in-plane
Zeeman fields on the MI-SF phase transition boundary and the nature
of the SF phase nearby. It is already established in the literature that
this method is equivalent to the mean-field decoupling theory at least
in the absence of a SOC, and therefore, it is expected to give
qualitatively accurate description of the MI and SF phases.
In addition to the phase diagrams, one of our main results is as follows:
Gutzwiller calculations showed that while the magnitudes of the order
parameters are uniform across the entire lattice, their phases may
vary from site to site due to SOC, and therefore, the SF phase is a
nonuniform one. We gave a complete account and intuitive understanding
of this SOC induced nonuniform-SF phase and its resultant phase
patterns, by supporting our numerical calculations with fully
analytical insights.

One may extend this work in many directions. For instance, as we
emphasized in the main text, recently proposed exotic magnetic
phases exhibiting spin textures in the form of spin spirals
and vortex and Skyrmion crystals inside the MI
lobes~\cite{cole12, radic12, cai12, zhou13, zhang03} are not accessible
within a variational Gutzwiller ansatz. However, there is no a priori
reason why the method can not be used to investigate the
spin-textured nonuniform-SF phases. Since spin-textured SF phases
have only been discussed in the literature for a weakly-interacting Bose gas,
such studies are especially of interest near the MI-SF phase
transition boundary and the strongly-interacting regime.
In addition, SF phases in this work are characterized by the order
parameter $\Delta_{j \alpha }=\langle \hat{a}_{j\alpha }\rangle$,
and therefore, single-particle and/or single-hole excitations are
always gapped inside our MI lobes. However, this definition of the
SF phase does not discriminate the possibility of exotic multi particle
and/or hole excitations that may be gapless, our MI lobes may still have
some sort of hidden SF orders. For instance, in the absence of SOC
and Zeeman field, it is already known that a counterflow-SF phase of
particle-hole pairs~\cite{kuklov03, altman03, hu09, ozaki12}
characterized by the order parameter
$\Delta _{j} \equiv \langle \hat{a}_{j\uparrow }\hat{a}_{j\downarrow}^{\dagger }\rangle$
and a paired SF phase of two particles or two
holes~\cite{arguelles07, altman03, hu09, iskin10, ozaki12}
characterized by the order parameter
$\Delta_{j} \equiv \langle \hat{a}_{j\uparrow }\hat{a}_{j\downarrow }\rangle$
are possible when $U_{\uparrow \downarrow} > 0$
and $U_{\uparrow \downarrow} < 0$, respectively~\cite{ozaki12}.
The effects of SOC and/or Zeeman field on the fates of such exotic SF
phases is an uncharted territory.

\section{Acknowledgments}

A. T. B is supported by T\"{U}B\.{I}TAK 2218 Domestic Postdoctoral Fellowship
Program, and M. I. is supported by the Marie Curie IRG Grant No.
FP7-PEOPLE-IRG-2010-268239, T\"{U}B\.{I}TAK Career Grant No. 3501-110T839,
and T\"{U}BA-GEB\.{I}P.

\bibliographystyle{apsrev4-1}

\end{document}